\documentclass[twocolappendix,appendixfloats,]{emulateapj}

\usepackage[colorlinks = true, linkcolor = blue, urlcolor  = blue, citecolor = blue, anchorcolor = blue]{hyperref}
\usepackage{hyperref}
\usepackage{apjfonts}
\usepackage{rotating}
\usepackage{amsmath}
\usepackage{color}
\usepackage{graphicx}
\usepackage[dvipsnames]{xcolor}
\usepackage{CJK}
\newcommand{\te}{t_{\rm E}}
\newcommand{\thetae}{\theta_{\rm E}}
\newcommand{\thetastar}{\theta_*}
\newcommand{\pie}{\pi_{\rm E}}

\newcommand{\dl}{D_{\rm L}}
\newcommand{\ds}{D_{\rm S}}

\definecolor{darkbrown}{RGB}{139,69,19}
\shorttitle{Wide-separation Microlensing Planet}
\shortauthors{Han et al.}

\begin{document}

%\title{OGLE-2016-BLG-1227L: Not a Free-floating Planet But a Wide-separation Microlensing Planet }
\title{OGLE-2016-BLG-1227L: A Wide-separation Planet from a Very Short-timescale Microlensing Event}

\author{
% leading author -----------------------------
Cheongho~Han$^{0001}$, 
Andrzej~Udalski$^{0003,100}$, 
Andrew~Gould$^{0004,0005,101}$
\\
(Leading authors),\\
and \\
% KMTNet ---------------------------
Michael~D.~Albrow$^{0007}$, 
Sun-Ju~Chung$^{0002,0008}$,  
Kyu-Ha~Hwang$^{0002}$, 
Youn~Kil~Jung$^{0002}$, 
Chung-Uk~Lee$^{0002,101}$, 
Yoon-Hyun~Ryu$^{0002}$, 
In-Gu~Shin$^{0002}$, 
Yossi~Shvartzvald$^{0009}$, 
Jennifer~C.~Yee$^{0010}$, 
Weicheng~Zang$^{0011}$,
Sang-Mok~Cha$^{0002,0012}$, 
Dong-Jin~Kim$^{0002}$, 
Hyoun-Woo~Kim$^{0002}$, 
Seung-Lee~Kim$^{0002,0008}$, 
Dong-Joo~Lee$^{0002}$, 
Yongseok~Lee$^{0002,0012}$, 
Byeong-Gon~Park$^{0002,0008}$, 
Richard~W.~Pogge$^{0005}$, 
M.~James~Jee$^{0013,0014}$, 
Doeon~Kim$^{0001}$,
Chun-Hwey~Kim$^{0016}$,
Woong-Tae~Kim$^{0017}$
\\
(The KMTNet Collaboration),\\
% OGLE -----------------------------
Przemek~Mr{\'o}z$^{0003,0018}$, 
Micha{\l}~K.~Szyma{\'n}ski$^{0003}$, 
Jan~Skowron$^{0003}$,
Radek~Poleski$^{0005}$, 
Igor~Soszy{\'n}ski$^{0003}$, 
Pawe{\l}~Pietrukowicz$^{0003}$,
Szymon~Koz{\l}owski$^{0003}$, 
Krzysztof~Ulaczyk$^{0015}$ \\
%Krzysztof~A.~Rybicki$^{0003}$,
%Patryk~Iwanek$^{0003}$ \\
%Marcin~Wrona$^{0003}$\\
(The OGLE Collaboration) \\   
}

%\email{cheongho@astroph.chungbuk.ac.kr}

%%=====================================
\affil{$^{0001}$ Department of Physics, Chungbuk National University, Cheongju 28644, Republic of Korea; cheongho@astroph.chungbuk.ac.kr} 
\affil{$^{0002}$ Korea Astronomy and Space Science Institute, Daejon 34055, Republic of Korea} 
\affil{$^{0003}$ Warsaw University Observatory, Al.~Ujazdowskie 4, 00-478 Warszawa, Poland} 
\affil{$^{0004}$ Max Planck Institute for Astronomy, K\"onigstuhl 17, D-69117 Heidelberg, Germany} 
\affil{$^{0005}$ Department of Astronomy, Ohio State University, 140 W. 18th Ave., Columbus, OH 43210, USA} 
\affil{$^{0007}$ University of Canterbury, Department of Physics and Astronomy, Private Bag 4800, Christchurch 8020, New Zealand} 
\affil{$^{0008}$ Korea University of Science and Technology, 217 Gajeong-ro, Yuseong-gu, Daejeon, 34113, Republic of Korea} 
\affil{$^{0009}$ Department of Particle Physics and Astrophysics, Weizmann Institute of Science, Rehovot 76100, Israel \color{black}}
\affil{$^{0010}$ Center for Astrophysics $|$ Harvard \& Smithsonian 60 Garden St., Cambridge, MA 02138, USA} 
\affil{$^{0011}$ Department of Astronomy and Tsinghua Centre for Astrophysics, Tsinghua University, Beijing 100084, China \color{black}} 
\affil{$^{0012}$ School of Space Research, Kyung Hee University, Yongin, Kyeonggi 17104, Republic of Korea} 
\affil{$^{0013}$ Yonsei University, Department of Astronomy, Seoul, Republic of Korea}
\affil{$^{0014}$ Department of Physics, University of California, Davis, California, USA}
\affil{$^{0015}$ Department of Physics, University of Warwick, Gibbet Hill Road, Coventry, CV4 7AL, UK} 
\affil{$^{0016}$ Department of Astronomy \& Space Science, Chungbuk National University, Cheongju 28644, Republic of Korea} 
\affil{$^{0017}$ Department of Physics \& Astronomy, Seoul National University, Seoul 08826, Republic of Korea}
\affil{$^{0018}$ Division of Physics, Mathematics, and Astronomy, California Institute of Technology, Pasadena, CA 91125, USA}
\altaffiltext{100}{OGLE Collaboration.}
\altaffiltext{101}{KMTNet Collaboration.}

\begin{abstract}
We present the analysis of the microlensing event OGLE-2016-BLG-1227. The light curve of 
this short-duration event appears to be a single-lens event affected by severe finite-source 
effects. Analysis of the light curve based on single-lens single-source (1L1S) modeling yields 
very small values of the event timescale, $\te\sim 3.5$~days, and the angular Einstein radius, 
$\thetae\sim 0.009$~mas, making the lens a candidate of a free-floating planet.  Close inspection 
reveals that the 1L1S solution leaves small residuals with amplitude $\Delta I\lesssim 0.03$~mag.  
We find that the residuals are explained by the existence of an additional widely-separated heavier 
lens component, indicating that the lens is a wide-separation planetary system rather than a 
free-floating planet.  From Bayesian analysis, it is estimated that the planet has a mass of 
$M_{\rm p} = 0.79^{+1.30}_{-0.39}~M_{\rm J}$ and it is orbiting a low-mass host star with a mass 
of $M_{\rm host}=0.10^{+0.17}_{-0.05}~M_\odot$ located with a projected separation of 
$a_\perp=3.4^{+2.1}_{-1.0}$~au.  The planetary system is located in the Galactic bulge with a 
line-of-sight separation from the source star of $D_{\rm LS}=1.21^{+0.96}_{-0.63}$~kpc.  The 
event shows that there are a range of deviations in the signatures of host stars for apparently 
isolated planetary lensing events and that it is possible to identify a host even when a deviation 
is subtle.
\end{abstract}


\section{Introduction}\label{sec:one}

Although most microlensing planets are detected through the channel of a short-term perturbation 
to the standard lensing light curve of the planet host \citep{Mao1991, Gould1992}, a fraction of 
planets can be detected through the channel of an isolated lensing event produced by the gravity 
of the planet itself \citep{Bennett2002, Han2004}.  The latter channel is important because it 
provides a unique method to probe free-floating planets (FFPs) that may have been ejected from 
the planetary systems in which they formed or have not been gravitationally bound to any host 
star before.

The most important characteristics of an FFP lensing event is its short timescale. This is because 
the event timescale $\te$ is related to the angular Einstein radius $\thetae$ and the relative 
lens-source proper motion $\mu$ by $\te=\thetae/\mu$, and the angular Einstein radius is proportional 
to the square root of the lens mass $M$, i.e.,
\begin{equation}
\thetae = (\kappa M \pi_{\rm rel})^{1/2},\qquad
\pi_{\rm rel}={\rm au}\left( {1\over \dl} - {1\over \ds} \right).
\label{eq1}
\end{equation}
Here $\kappa=4G/(c^2{\rm au})$, $\pi_{\rm rel}$ represents the relative lens-source parallax, and 
$\dl$ and $\ds$ denote the distances to the lens and source, respectively.  For FFP events, the 
chance to exhibit deformed lensing light curves caused by severe finite-source effects is high.  
The deformation can occur when the angular source radius $\thetastar$ is comparable to $\thetae$ 
for FFP event, in which case the light curve is very likely to be affected by severe finite-source 
effects.  Recently, three candidates of FFPs were reported by \citet{Mroz2018} and \citet{Mroz2019} 
from the analyses of the lensing events with these characteristics.

However, even when an event is both very short and exhibits strong finite-source effects, the 
lens cannot be securely identified as an ``FFP''.  First, it is always possible that the small 
value of $\thetae$ derives from a small $\pi_{\rm rel}$ rather than small lens mass $M$.  See 
Equation~(\ref{eq1}). This issue can only be resolved for individual FFP candidates by measuring 
the microlens parallax $\pie\equiv \pi_{\rm rel}/\thetae$, using, e.g., a satellite in solar orbit 
\citep{Refsdal1966} or so-called terrestrial parallax \citep{Gould1997,Gould2009}. Nevertheless, 
from an ensemble of $\thetae$ measurements (even without corresponding $\pie$ measurements), one 
can statistically constrain the properties of the FFP population.  However, there is a second 
fundamental problem that has the potential to corrupt such a statistical sample, namely that a 
wide-separation planet can also produce a lensing light curve with similar characteristics, 
masquerading as an FFP.  Therefore, it is important to distinguish the two populations of events 
produced by FFPs and wide-separation planets in order to draw statistically meaningful conclusions 
about the properties and frequency of both bound and unbound planets.

% Figure 1 ------------------------------------------------------
\begin{figure}
\includegraphics[width=\columnwidth]{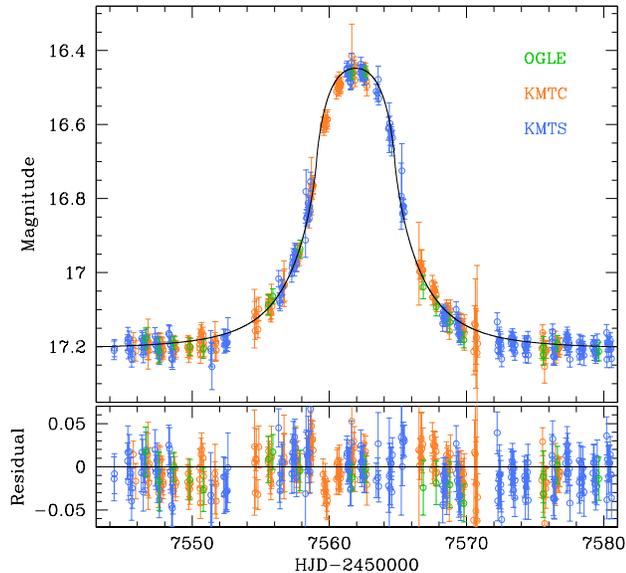}
\caption{
Lightcurve of OGLE-2016-BLG-1227. The curve drawn on the data points is the model 
obtained from the 1L1S fitting to the light curve considering finite-source effects.
\smallskip
}
\label{fig:one}
\end{figure}
% --------------------------------------------------------------

\citet{Han2003a} pointed out that an important fraction of isolated short-timescale events 
produced by wide-separation planets can be distinguished from those produced by FFPs
by detecting the signatures of host stars in the lensing light curves.\footnote{Besides 
this method, the nature of a wide-separation planet can be identified by several other methods. 
One method is detecting long-term bumps in the light curve caused by the primary star \citep{Han2005}. 
Another method is detecting the blended light from a host star by conducting high-resolution observations 
\citep{Bennett2002}. The last proposed method is conducting astrometric follow-up observations 
of isolated events using high-precision interferometers \citep{Han2006b}.} The signatures arise due 
to the planetary caustic induced by the binarity of the planet-host system.  For a binary lens 
composed of a planet and a host, there exist two sets of caustics. One set of caustics is located 
close to the host (central caustic) and the other caustic (planetary caustic) is located at a 
distance of $s_{\rm c}=s-1/s$ from the host. Here $s$ represents the projected planet-host 
separation normalized to $\thetae$. The planetary caustic of a wide-separation planet forms a 
closed curve with 4 cusps.  The full width along the star-planet axis, $\Delta\xi_{\rm c}$, 
and the height normal to the star-planet axis, $\Delta\eta_{\rm c}$, of the caustic are
\begin{equation}
\Delta\xi_{\rm c}  = {4q^{1/2}\over s \sqrt{s^2-1}};\qquad
\Delta\eta_{\rm c} = {4q^{1/2}\over s \sqrt{s^2+1}}
\label{eq2}
\end{equation}
respectively \citep{Han2006a}. For a wide-separation planet with $s\gg 1$, the planetary caustic 
is located close to the planet, i.e., $s_{\rm c}\rightarrow s$, and both $\Delta\xi_{\rm c}$ and 
$\Delta\eta_{\rm c}$ approaches $4q^{1/2}s^{-2}$, forming an astroid-shape caustic. The caustic 
size rapidly shrinks with the increase of the planet-host separation, i.e., 
$\Delta\xi_{\rm c}\sim \Delta\eta_{\rm c} \propto s^{-2}$.  As the caustic becomes smaller, the 
signature of the host star diminishes with the increasing finite-source effects.

In this paper, we present the analysis of the lensing event OGLE-2016-BLG-1227. The light curve 
of the event appears to be approximated by a short-timescale 1L1S model with severe finite-source 
effects, making the lens a candidate FFP. From the close inspection of the light curve, 
it is found that the 1L1S solution leaves small residuals.  We inspect the origin of the residuals 
to check the existence of a widely-separated heavier lens component, i.e., host of the planet.

We organize the paper as follows.  In Section~\ref{sec:two}, we describe the observations of the 
lensing event and the data obtained from these observations.  In Section~\ref{sec:three}, we present 
the analysis of the event based on the 1L1S interpretation.  In Section~\ref{sec:four}, we inspect 
the possible existence of a widely separated host of the planet by conducting a binary-lens (2L1S) 
analysis.  In Section~\ref{sec:five}, we estimate the angular Einstein radius by determining the 
dereddened color and brightness of the source star.  In Section~\ref{sec:six}, we conduct Bayesian 
analysis of the event to determine the physical lens parameters including the mass and location of 
the lens system.  We summarize the results and conclude in Section~\ref{sec:seven}.

% Figure 2 ------------------------------------------------------
\begin{figure}
\includegraphics[width=\columnwidth]{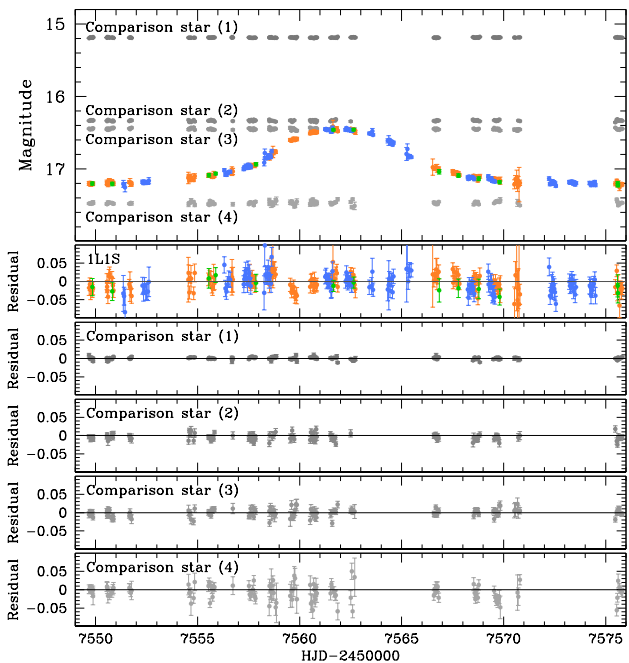}
\caption{
Comparison of the lensing lightcurve with those of four comparison stars around the lensing source.  
The lower four panels show the residuals of the comparison stars from baseline magnitudes and the 
second panel shows the residuals of the lensing event from the 1L1S solution.
\smallskip
}
\label{fig:two}
\end{figure}
% --------------------------------------------------------------

% Table 1 ------------------------------------------------
\begin{deluxetable}{lcc}
\tablecaption{Data Used in Analysis\label{table:one}}
\tablewidth{240pt}
%\tabletypesize{\small}
\tablehead{
\multicolumn{1}{c}{Data set}          &
\multicolumn{1}{c}{$N_{\rm data}$}    &
\multicolumn{1}{c}{Range (${\rm HJD}^\prime$)}           
}
\startdata                                              
OGLE & 154   & 7110.8 -- 7659.6   \\
KMTC & 369   & 7500.7 -- 7599.7   \\
KMTS & 575   & 7441.6 -- 7675.3   
\enddata                            
\tablecomments{
$N_{\rm data}$ indicates the number of each data set.
\smallskip
}
\end{deluxetable}
\bigskip
% --------------------------------------------------------------

% Table 2 ------------------------------------------------
\begin{deluxetable*}{lccc}
\tablecaption{Lensing parameters\label{table:two}}
\tablewidth{480pt}
%\tabletypesize{\small}
\tablehead{
\multicolumn{1}{c}{Parameter}       &
\multicolumn{1}{c}{1L1S}            &
\multicolumn{2}{c}{2L1S}            \\
\multicolumn{1}{c}{}                &
\multicolumn{1}{c}{}                &
\multicolumn{1}{c}{Inner solution}  &
\multicolumn{1}{c}{Outer solution}           
}
\startdata                                              
$\chi^2$                          &  1115.1                 &  968.6                 &   973.0                   \\
$t_0$ (${\rm HJD}^\prime$)        &  7561.920 $\pm$ 0.017   &  7561.999 $\pm$ 0.031  &   7561.976 $\pm$ 0.032    \\
$u_0$                             &  0.681 $\pm$ 0.017      &  0.066 $\pm$ 0.012     &   $-0.057 \pm 0.012$      \\
$t_{\rm E}$   (days)              &  3.54 $\pm$ 0.05        &  45.37 $\pm$ 8.07      &   52.23 $\pm$ 12.76       \\
$t_{\rm E,1}$ (days)              &  --                     &  4.05 $\pm$ 0.06       &   4.01 $\pm$ 0.06         \\
$t_{\rm E,2}$ (days)              &  --                     &  45.19 $\pm$ 8.09      &   52.07 $\pm$ 12.79       \\
$s$                               &  --                     &  3.68 $\pm$ 0.21       &   3.57 $\pm$ 0.24         \\
$q$                               &  --                     &  124.48 $\pm$ 46.79    &   168.99 $\pm$ 98.86      \\
$\alpha$ (rad)                    &  --                     &  4.783 $\pm$ 0.062     &   4.689 $\pm$ 0.066       \\
$\rho$                            &  1.05 $\pm$ 0.013       &  0.092 $\pm$ 0.017     &   0.080 $\pm$ 0.018       \\
% ----------------
$t_{\rm eff}=|u_0| \te$   (days)  & --                      &  3.00 $\pm$ 0.08        &  2.97 $\pm$ 0.08         \\  
$t_*=\rho\te$           (days)    & --                      &  4.17 $\pm$ 0.03        &  4.16 $\pm$ 0.03         \\  
$t_{\rm p}=q^{-1/2}\te$ (days)    & --                      &  4.07 $\pm$ 0.06        &  4.02 $\pm$ 0.06           
\enddata                            
\tablecomments{
${\rm HJD}^\prime = {\rm HJD}- 2450000$. For the 2L1S solution, $\te$ represents the event timescale
corresponding to the total mass of the binary lens, and $t_{\rm E,1}$ and $t_{\rm E,2}$ represent the
timescales corresponding to the masses of individual lens components, $M_1$ and $M_2$, respectively. 
The subscripts of the lens components are chosen according to the distances from the source
trajectory. The source trajectory passes closer to the lower-mass lens component and thus
$M_1<M_2$, $t_{\rm E,1}<t_{\rm E,2}$, and $q=M_2/M_1>1$.
\smallskip
}
\end{deluxetable*}

\section{Observation and Data}\label{sec:two}

The lensing event OGLE-2016-BLG-1227 occurred on a star located toward the Galactic bulge 
field.  The equatorial coordinates of the lensed star (source) are 
$({\rm R.A.},{\rm decl.})_{\rm J2000}=(17:42:23.31, -33:45:35.2)$, which correspond to 
the galactic coordinates $(l,b)=(-4^\circ\hskip-2pt.47, -1^\circ\hskip-2pt.94)$. The source 
of the event is a bright giant with a baseline magnitude of $I_{\rm base}=16.89$ from the 
calibrated OGLE photometric maps.

The lensing event was first discovered by the Optical Gravitational Lensing Experiment
\citep[OGLE:][]{Udalski2015} survey, and the discovery was notified to the microlensing community 
on 2016 June 29. The OGLE survey was conducted utilizing the 1.3~m telescope located at the Las 
Campanas Observatory in Chile.  The telescope is equipped with a camera, which consists of 32 
$2{\rm k}\times 4{\rm k}$ chips, yielding a 1.4~deg$^2$ field of view.  The OGLE images were 
obtained mostly in $I$ band and some images were taken in $V$ band for the source color measurement.

The event was also located in the field toward which the Korea Microlensing Telescope Network 
survey \citep[KMTNet:][]{Kim2016} was monitoring.  The KMTNet survey was conducted using the
three identical 1.6~m telescopes that are globally distributed in the southern hemisphere at 
the Siding Spring Observatory in Australia (KMTA), Cerro Tololo Interamerican Observatory in Chile 
(KMTC), and the South African Astronomical Observatory in South Africa (KMTS). Each KMTNet telescope 
is equipped with a camera, consisting of four $9{\rm k}\times 9{\rm k}$ chips, yielding 4~deg$^2$ 
field of view.  The event was found from the analysis of the data conducted after the 2016 season 
\citep{Kim2018} and it was designated as KMT-2016-BLG-1089.  Most KMTNet images were obtained in 
$I$ band and about one tenth of images were obtained in $V$ band for the source color measurement. 
Thanks to the high-cadence coverage ($1~{\rm hr}^{-1}$ for each telescope) using the multiple 
telescopes, the detailed structure of the light curve is well delineated by the KMTNet data, 
despite the short duration of the event.

% Figure 3 ------------------------------------------------------
\begin{figure*}
\epsscale{0.90}
\plotone{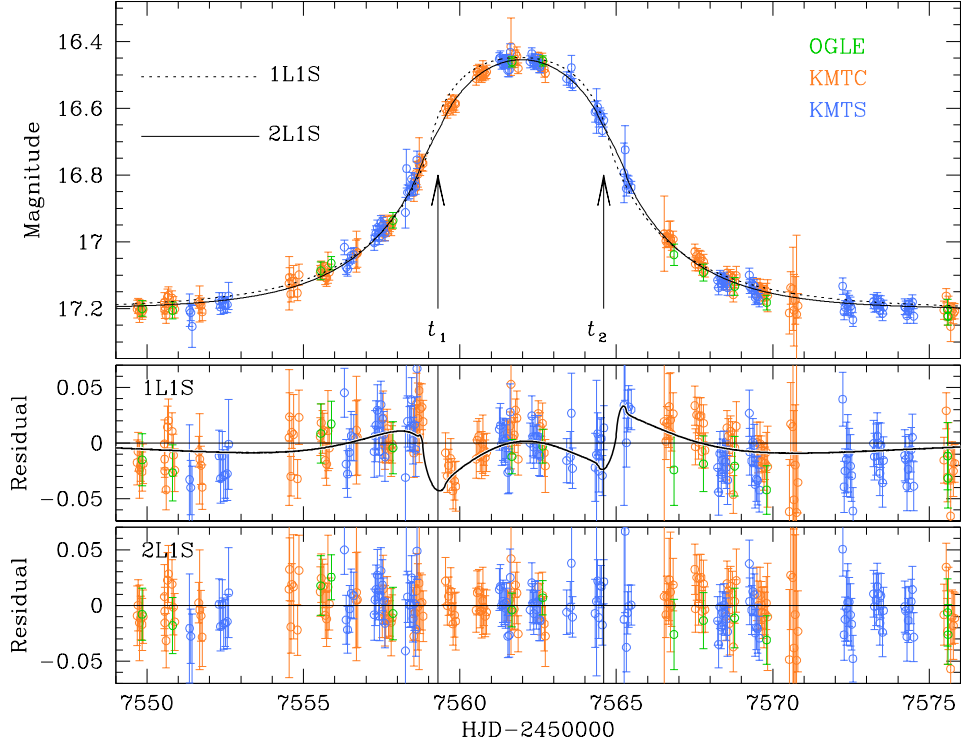}
\caption{
Comparison of the 1L1S (dotted curve) and 2L1S (solid curve) solutions.  The middle and 
bottom panels show the residuals from the 1L1S and 2L1S solutions, respectively.  The 
solid curve in the middle panel represents the difference between the 1L1S and 2L1S solutions.  
In the top panel, the arrows at $t_1({\rm HJD}^\prime)=7559.3$ and 
$t_2({\rm HJD}^\prime)=7564.6$ represent the times of the two dips in the residuals from 
the 1L1S solution.
\bigskip
}
\label{fig:three}
\end{figure*}
% --------------------------------------------------------------

Reduction of the data was carried out using the photometry codes developed by the individual survey 
groups: \citet{Wozniak2000} for the OGLE and \citet{Albrow2009} for the KMTNet data sets. 
These codes are based on the difference imaging method developed by \citet{Alard1998}. For a subset 
of the KMTNet data sets, additional photometry is conducted using the pyDIA code \citep{Albrow2017} 
to measure the source color. The errorbars of the individual data sets are readjusted according to 
the procedure described in \citet{Yee2012}.  We note that the KMTA data set is not used in the analysis 
because the photometry quality is relatively low and the data do not cover the major part of the light 
curve.  In Table~\ref{table:one}, we list the data sets used in the analysis along with numbers of 
data points, $N_{\rm data}$, and the time ranges of the individual data sets.

\section{Single-lens single-source (1L1S) modeling}\label{sec:three}

In Figure~\ref{fig:one}, we present the light curve of OGLE-2016-BLG-1227.  The light curve appears 
to be that of a 1L1S event affected by severe finite-source effects. We, therefore, start 
the analysis of the event by conducting a 1L1S modeling.

The modeling is carried out by searching for the lensing parameters that best describe the observed 
light curve. The light curve of a 1L1S event affected by finite-source effects is described by four 
lensing parameters. These parameters include the time of the closest lens-source approach, $t_0$, 
the lens-source separation at that time, $u_0$, the event timescale, $t_{\rm E}$, and the normalized 
source radius, $\rho$.  The normalized source radius is defined as the ratio of the angular source 
radius $\theta_*$ to the angular Einstein radius, i.e., $\rho=\theta_*/\thetae$, and it is needed 
to describe the deformed light curve caused by finite-source effects.  We search for the best-fit 
lensing parameters using the Markov Chain Monte Carlo (MCMC) method.

In computing finite-source magnifications, we consider the variation of the source surface 
brightness caused by limb darkening \citep{Witt1995, Valls1995, Loeb1995}.  To account for the 
limb-darkening variation, we model the surface brightness of the source star as
\begin{equation}
S_\lambda = \bar{S}_\lambda 
\left[ 1-\Gamma_\lambda \left( 1-{3\over 2}\cos\phi \right)\right],
\label{eq3}
\end{equation}
where $\bar{S}_\lambda$ denotes the mean surface brightness, $\Gamma_\lambda$ is the linear 
limb-darkening coefficient, and $\phi$ represents the angle between the line of sight toward the 
center of the source star and the normal to the source surface. The limb-darkening coefficient is 
determined based on the stellar type of the source star. As we will show in Section~\ref{sec:five}, 
the source is a bulge giant with a spectral type K3. Based on the stellar type, we set the limb-darkening 
coefficient as $\Gamma_I = 0.41$ and $\Gamma_V = 0.74$ by adopting the values from \citet{Claret2000} 
under the assumption that $v_{\rm turb}=2~{\rm km}~{\rm s}^{-1}$, $\log(g/g_\odot)=-2.4$, and 
$T_{\rm eff}=4500~{\rm K}$.  For the computation of finite-source magnifications, we use the 
semianalytic expressions derived by \citet{Gould1994} and \citet{Witt1994}.

In Table~\ref{table:two}, we present the best-fit lensing parameters obtained from the 1L1S modeling.
In Figure~\ref{fig:one}, we also present the model curve superposed on the data points.  We note 
that the estimated event timescale, $\te\sim 3.5$~days, is much shorter than those of typical lensing 
events with $\sim (O)10$~days although events with such short timescales are not extremely rare.  
Furthermore, the normalized source radius, $\rho\sim 1.05$, is much bigger than typical values of 
$\sim 0.01$ -- 0.02 for events involved with giant source stars.  The unusually large $\rho$ value 
suggests that the angular Einstein radius is likely to be very small.  As we will show in 
Section~\ref{sec:five}, the angular radius of the source is $\theta_*\sim 9.0~\mu{\rm as}$, and thus 
the angular Einstein radius of the event is $\thetae\sim 0.009$~mas. This is very much smaller than 
$\sim 0.5$~mas of typical lensing events. The very small values of $\te$ and $\thetae$ make the lens 
of the event a candidate of an FFP or a brown-dwarf.  We note that the lens of the event was originally 
found as a brown-dwarf or an FFP candidate from the search for isolated events with short $\te$ and 
very small $\thetae$ conducted by \citet{Han2019}, but the analysis is separately presented in this 
work for the reason presented in Section~\ref{sec:four}.

Although the observed light curve appears to be approximated by the 1L1S model, it is found that 
the solution leaves small residuals with amplitude $\Delta I\lesssim 0.03$~mag.  See the lower 
panel of Figure~\ref{fig:one}.  The source was located close to the Moon during the lensing 
magnification and thus the photometry We check this possibility by conducting additional photometry 
for nearby stars.  In Figure~\ref{fig:two}, we present the lightcurves of four comparison stars and 
compare them with that of the lensing event.  It shows that the magnitudes of the comparison stars 
remain constant in contrast to the 1L1S residuals.  This indicates that the photometry is not 
affected by the Moon and the residuals from the 1L1S solution are likely to be real.

\section{Binary-lens single-source (2L1S) modeling}\label{sec:four}

Considering that the main part of the lensing light curve is 
produced by a planetary-mass object, we check whether there exists a host star located away from 
the planet. For this, we additionally conduct a 2L1S modeling of the light curve.

Compared to the 1L1S modeling, the 2L1S modeling requires three additional lensing parameters to 
describe the lens binarity.  These parameters include the projected binary separation normalized 
to the angular Einstein radius, $s$, the mass ratio between the lens components, $q=M_2/M_1$, and 
the angle between the binary axis and the source trajectory, $\alpha$ (source trajectory angle).

In the 2L1S modeling, the solution of the lensing parameters is searched for in two steps. In 
the first step, we conduct a grid search for the parameters $s$ and $q$, while the other parameters 
are searched for using the MCMC method. This procedure yields a $\chi^2$ map on the $s$--$q$ 
parameter plane and we find local minima that appear in the map. In the second step, we refine the 
individual local minima by additionally conducting modeling with all parameters, including the grid 
parameters $s$ and $q$, allowed to vary.  We find a global solution by comparing the goodness of 
the local solutions.  This procedure allows us to find degenerate solutions, if they exist.

% Figure 4 ------------------------------------------------------
\begin{figure}
\includegraphics[width=\columnwidth]{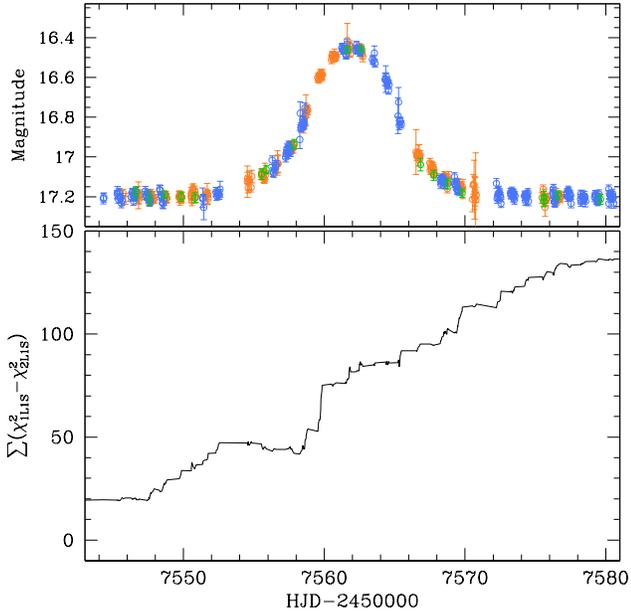}
\caption{
Cumulative distribution of $\Delta\chi^2$ between the 1L1S and 2L1S models (lower panel). 
The light curve in the upper panel is presented to show the region of fit improvement.
\smallskip
}
\label{fig:four}
\end{figure}
% --------------------------------------------------------------

We find that the model fit substantially improves with the introduction of an additional widely-separated 
lens component $M_2$.  The additional lens component has a mass much heavier than the lens component 
$M_1$ responsible for the short magnified part of the light curve, suggesting that the additional lens 
component is the host of the planet. In Figure~\ref{fig:three}, we present both the 1L1S and 2L1S models 
and the residuals from the individual models. The solid curve superposed on the residuals of the 1L1S 
model in the middle panel represents the difference between the 1L1S and 2L1S models. It is found that 
the 2L1S residuals are substantially reduced relative to the 1L1S model.  In Figure~\ref{fig:four}, 
we present the cumulative distribution of $\Delta\chi^2=\chi^2_{\rm 1L1S}-\chi^2_{\rm 2L1S}$ between 
the 1L1S and 2L1S models to better show the region of the fit improvement. We find that the 2L1S 
improves the fit by $\Delta\chi^2\sim 146.5$. We further check whether there is an additional weak 
long-term bump caused by the heavier companion, but we find no such a bump.  As we will show below, 
the reason for the absence of a bump is that the source passes perpendicular to the binary axis.

In searching for lensing solutions, we find that the observed light curve is subject to the 
so-called ``inner/outer 
degeneracy''.  This degeneracy arises because the planetary anomalies produced by the source approaching 
the inner and outer sides (with respect to the host of the planet) of the planetary caustic are similar 
to each other \citep{Gaudi1997}.  It is found that the degeneracy is severe although the inner solution 
is slightly preferred over the outer solution by $\Delta\chi^2\sim 4.3$.

In Table~\ref{table:two}, we list the best-fit lensing parameters of the 2L1S solutions for both the 
inner and outer solutions.  For each solution, we present three values of timescales 
($\te$, $t_{\rm E,1}$, $t_{\rm E,2}$), in which $\te$ 
represents the event timescale corresponding to the total mass of the binary lens, and 
$t_{\rm E,1}=[1/(1+q)]^{1/2}\te$ and $t_{\rm E,2}=[q/(1+q)]^{1/2}\te$ represent the timescales 
corresponding to the masses of individual lens components, $M_1$ and $M_2$. We note that the subscripts 
of the lens components $M_1$ and $M_2$ are chosen according to the distances from the source trajectory. 
The source trajectory approaches closer to the lower-mass lens component and thus $M_1<M_2$, 
$t_{\rm E,1}<t_{\rm E,2}$, and $q=M_2/M_1>1$. The estimated mass ratio between the lens components, 
$q\sim 124$ for the inner solution and $q\sim 169$ for the outer solution, is much bigger than unity, 
indicating that $M_2$ is the host of the planet $M_1$. The host is separated from the planet with a 
projected separation of $s\sim 3.6$.

% Figure 5 ------------------------------------------------------
\begin{figure}
\includegraphics[width=\columnwidth]{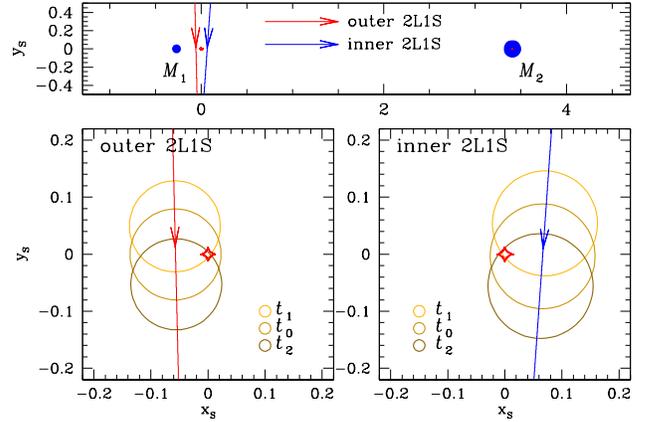}
\caption{
Lens-system configurations of the inner and outer 2L1S solutions. Coordinates are centered at 
the center of the planetary caustic. The time $t_0$ is the time of the closest source approach 
to the planetary caustic, and the times $t_1$ and $t_2$ correspond to the times of the two dips 
in the residuals from the 1L1S model presented in Fig.~\ref{fig:four}. In each panel, the line with 
an arrow represents the source trajectory. The circles on the source trajectory in the two lower 
panels represent the source positions at $t_0$, $t_1$, and $t_2$. The size of the circle is scaled 
to the source size.
\bigskip
}
\label{fig:five}
\end{figure}
% --------------------------------------------------------------

In Figure~\ref{fig:five}, we present the lens-system configurations of the inner and outer 2L1S
solutions. The upper panel shows the whole view including the both lens components.  The lower 
two panels show the zoom of the region around the planetary caustic for the inner (right panel) 
and outer (left panel) solutions. The three brown-tone circles in the lower panels represent 
the source positions at three different times of $t_0$, $t_1$, and $t_2$. The time $t_0$ 
corresponds to the time of the closest source approach to the planetary caustic, and the times 
$t_1({\rm HJD}^\prime)=7559.3$ and $t_2({\rm HJD}^\prime)=7564.6$ correspond to the times of 
the two dips in the residuals from the 1L1S model.  See the corresponding times $t_1$ and $t_2$ 
marked in Figure~\ref{fig:three}.  The size of the circles is scaled to the source size. It is 
found that the source is much bigger than the caustic. This causes severe attenuation of the 
signal induced by the caustic and makes the light curve appear to be very similar to that of a 
1L1S event.

We note that the estimated lensing parameters have large uncertainties.  See Table~\ref{table:two}. 
The main reason for the large uncertainties of the lensing parameters is that the observed 
lensing magnification is mostly produced by the planet, and the planet's host is characterized 
by the subtle deviations in the planet-induced magnifications.  In this case, the uncertainty 
of the timescale $t_{\rm E}\sim t_{\rm E,2}$ is large.  The large uncertainty of $\te$ 
propagates into the uncertainty of mass ratio because the mass ratio is related to the timescale 
by $q=(t_{\rm E,1}/t_{\rm E,2})^{1/2}\sim (t_{\rm E,1}/t_{\rm E})^{1/2}$.  The uncertain timescale 
also induces large uncertainties of $u_0$ and $\rho$ because the measured caustic-crossing duration 
results from the combination of these parameters by $t_{\rm cc}  = 2(u_0^2+\rho^2)^{1/2}\te$.

% Figure 6 ------------------------------------------------------
\begin{figure}
\includegraphics[width=\columnwidth]{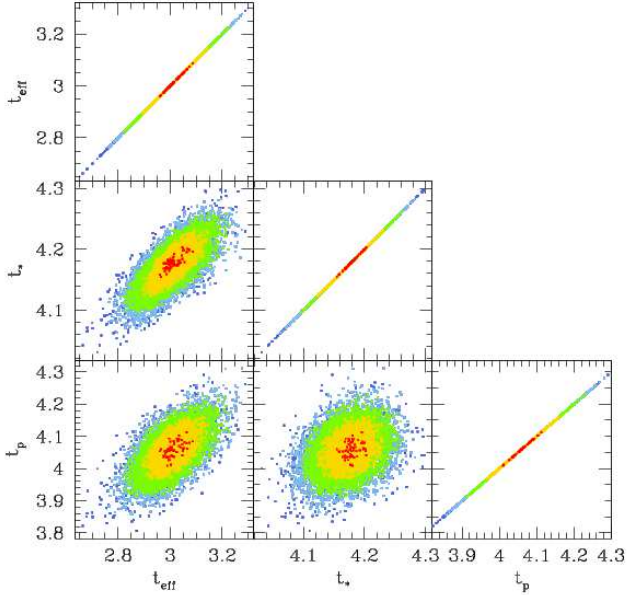}
\caption{
$\Delta\chi^2$ distributions of points in the MCMC chain  
on the parameter planes of the $(t_{\rm eff}, t_*, t_{\rm p})$ combinations.
The red, yellow, green, and blue colors represent points with 
$1\sigma$, $2\sigma$, $3\sigma$, and $4\sigma$, respectively.
The distributions are constructed based on the ``inner 2L1S solution''. 
\smallskip
}
\label{fig:six}
\end{figure}
% --------------------------------------------------------------

In Figure~\ref{fig:six}, we present the $\Delta\chi^2$ distributions of points in the MCMC chain
on the $t_{\rm eff}$--$t_*$--$t_{\rm p}$ parameter planes.  The individual timescales represent 
$t_{\rm eff}=|u_0| \te$, $t_* = \rho \te$, and $t_{\rm p} = q^{-1/2}\te$, respectively.  The 
``effective timescale'' $t_{\rm eff}$ is frequently used because it facilitates intuitive 
understanding of a light curve independent of separately determining $u_0$ and $\te$ from modeling.  
The ``source-crossing timescale'' $t_*$ represents an approximate timescale for the lens to 
transit the source surface.  Finally, the ``planet timescale'' $t_{\rm p}$ denotes an approximate 
timescale of the isolated event produced by the planet.  We present the estimated values of these 
timescales in Table~\ref{table:two}.  These timescales are derived from the shape of a lensing light 
curve, and thus they are tightly constrained despite the large uncertainties of the lensing parameters, 
as demonstrated in Figure~\ref{fig:six}.

\section{Angular Einstein radius}\label{sec:five}

We determine the angular Einstein radius from the normalized source radius $\rho$ together 
with the angular source radius $\theta_*$ by $\thetae=\theta_*/\rho$. The normalized 
source radius is determined from modeling the light curve. For the estimation of the 
angular source radius, we use the method of \citet{Yoo2004}.  According to this method, 
we first place the source position in the instrumental color-magnitude diagram (CMD) of 
stars around the source. We then measure the offsets in color, $\Delta (V-I)$, and magnitude, 
$\Delta I$, of the source from the centroid of the red giant clump (RGC) in the CMD. With the 
measured offsets $\Delta (V-I)$ and $\Delta I$ together the known dereddened source color 
and magnitude of the RGC centroid, $(V-I,I)_{\rm RGC,0}= (1.06,14.65)$ \citep{Bensby2013, Nataf2013}, 
the dereddened color and magnitude of the source are estimated by
\begin{equation}
(V-I,I)_0=(V-I,I)_{\rm RGC,0}+\Delta (V-I,I).
\label{eq4}
\end{equation}

% Figure 7 ------------------------------------------------------
\begin{figure}
\includegraphics[width=\columnwidth]{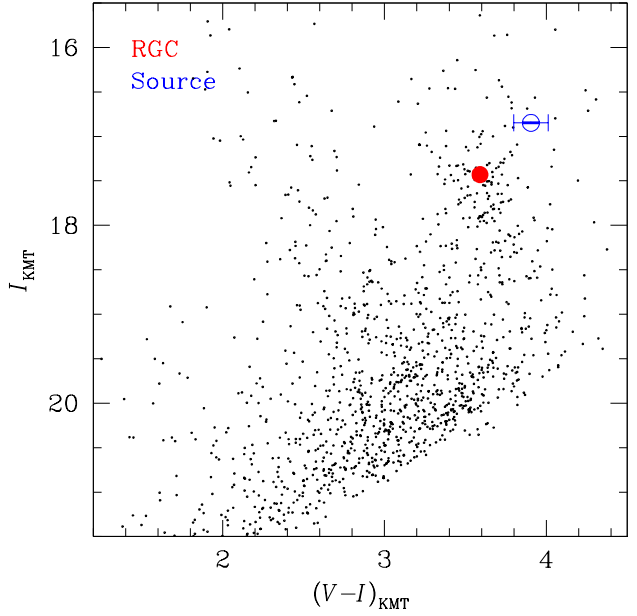}
\caption{
Source location (blue empty circle) with respect to the centroid of red
giant clump (RGC, red dots) in the instrumental color-magnitude diagram
constructed based on the pyDIA photometry of the KMTC data set.
\smallskip
}
\label{fig:seven}
\end{figure}
% --------------------------------------------------------------

In Figure~\ref{fig:seven}, we present the positions of the source and the RGC centroid in the 
instrumental CMD. The CMD is constructed using the pyDIA photometry of the KMTC data set. We 
note that the location of the blend cannot be determined because the baseline flux is dominated 
by the source flux and the flux from the blend is consistent with zero within the photometry 
uncertainty. The color and magnitude of the source in the instrumental CMD are 
$(V-I, I)=(3.91\pm 0.11, 16.85\pm 0.01)$ compared to those of the RGC centroid of 
$(V-I, I)_{\rm RGC}=(3.59, 17.43)$. With the measured offsets of $\Delta (V-I)=0.32\pm 0.11$ and 
$\Delta I=0.58\pm 0.01$, the de-reddened color and brightness of the source are estimated as 
$(V-I, I)_0 = (1.38\pm 0.11, 14.07\pm 0.01)$. The estimated source color and brightness indicate 
that the source is a typical bulge giant with a spectral type K3.

Once the dereddened color and magnitude are determined, we then estimated the angular source 
radius. For this, we first convert the $V-I$ color into $V-K$ color using the color-color relation 
of \citet{Bessell1988} and then the angular source radius is estimated using the \citet{Kervella2004} 
relation between $V-K$ and $\theta_*$. This procedure yields an angular source radius of
\begin{equation}
\theta_* = 9.01 \pm 1.15~\mu{\rm as}. 
\label{eq5}
\end{equation}
With the measured angular source radius, the angular Einstein radius is estimated as
\begin{equation}
\thetae= {\theta_*\over \rho}=
\begin{cases}
0.098\pm 0.044~{\rm mas} & \text{(inner solution)}, \\
0.113\pm 0.058~{\rm mas} & \text{(outer solution)},
\end{cases}
\label{eq6}
\end{equation}
The estimated relative lens-source proper motion is 
\begin{equation}
\mu= {\thetae\over \te}= 0.79\pm 0.10~{\rm mas~yr}^{-1}
\label{eq7}
\end{equation}
for both the inner and outer solutions.  We note that the fractional uncertainty of the relative 
lens-source proper motion, $\sigma_\mu/\mu\sim 13\%$, is substantially smaller than the uncertainty 
of the angular Einstein radius, $\sigma_{\thetae}/\thetae\sim 50\%$.  This is because the proper 
motion in the lensing modeling is computed by $\mu \sim \theta_*/t_*$ and the uncertainty of the 
``source-crossing timescale'' $t_*$ is significantly smaller than the uncertainty of the event 
timescale $\te$.

% Table 3 ------------------------------------------------
\begin{deluxetable}{lcc}
\tablecaption{Angular Einstein radius and relative lens-source proper motion\label{table:three}}
\tablewidth{240pt}
%\tabletypesize{\small}
\tablehead{
\multicolumn{1}{c}{Parameter}       &
\multicolumn{1}{c}{Inner solution}  &
\multicolumn{1}{c}{Outer solution}           
}
\startdata                                              
$\thetae$ (mas)            &  0.098 $\pm$ 0.044      &  0.113 $\pm$ 0.058              \\
$\theta_{\rm E,1}$ (mas)   &  0.007 $\pm$ 0.003      &  0.009 $\pm$ 0.004              \\
$\theta_{\rm E,2}$ (mas)   &  0.097 $\pm$ 0.043      &  0.112 $\pm$ 0.057              \\
$\mu$ (mas~yr$^{-1}$)      &  0.79 $\pm$ 0.10        &  0.79 $\pm$ 0.10              
\enddata                            
\tablecomments{
The Einstein radius $\thetae$ corresponds to the total mass of the lens $M=M_1+M_2$, and
$\theta_{\rm E,1}$ and $\theta_{\rm E,2}$ represent the Einstein radii corresponding to 
$M_1$ and $M_2$, respectively.
\smallskip
}
\end{deluxetable}
\bigskip
% --------------------------------------------------------------

In Table~\ref{table:three}, we summarize the estimated Einstein radii and relative lens-source 
proper motions for the inner and outer solutions. Also presented are the angular Einstein 
radii corresponding to the masses of the individual lens components, $\theta_{\rm E,1}$, and 
$\theta_{\rm E,2}$, similar to the presentation of $t_{\rm E,1}$ and $t_{\rm E,2}$ in 
Table~\ref{table:two}.  We note that the estimated $\theta_{\rm E,1}\sim 0.007$ -- 0.009~mas 
is consistent with the Einstein radius estimated from the 1L1S modeling. We also note that the 
measured angular Einstein radius, $\thetae \sim 0.1$~mas, is substantially smaller than 
$\sim 0.5$~mas of a typical lensing event produced by a low-mass star with a mass of 
$\sim 0.3~M_\odot$ located roughly halfway between the observer and the bulge source. The 
angular Einstein radius is related to the lens mass and distance by Equation~(\ref{eq1}).  
Then, the small angular Einstein radius suggests that the lens has a small mass and/or it 
is located close to the source.

\section{Physical lens parameters}\label{sec:six}

For the unique determinations of the physical lens parameters of the lens mass $M$ and distance $\dl$, 
one must measure both the angular Einstein radius and the microlens parallax $\pie$, i.e.,
\begin{equation}
M={\thetae\over \kappa\pie};\qquad
\dl = {{\rm au} \over \pie\thetae + \pi_{\rm S}}.
\label{eq8}
\end{equation}
Here $\pi_{\rm S}={\rm au}/D_{\rm S}$ represents the parallax of the source. For OGLE-2016-BLG-1227, 
the angular Einstein radius is measured from the obvious finite-source effects, but the microlens 
parallax cannot be measured due to the short timescale of the observed light curve, i.e., $t_{\rm E,1}$.  
We, therefore, estimate $M$ and $\dl$ by conducting Bayesian analysis of the event based on the measured 
event timescale $\te$ and   the relative lens-source proper motion $\mu$.  We use $\mu$ instead of 
$\thetae$ because $\te$ and $\thetae$ are highly correlated.

% Figure 8 ------------------------------------------------------
\begin{figure}
\includegraphics[width=\columnwidth]{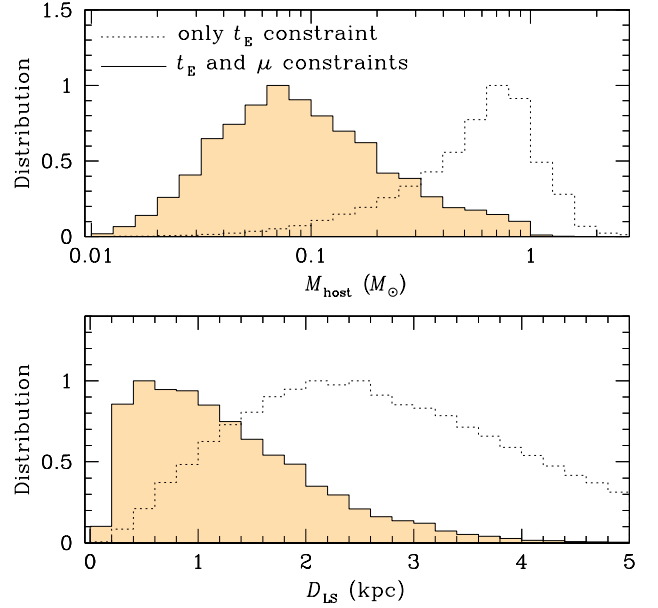}
\caption{
Probability distributions of the lens mass of the planet host ($M_{\rm host}$) and the lens-source 
separation ($D_{\rm LS}$) obtained from the Bayesian analysis.  The solid curve curve is the distribution 
obtained with the combined $\thetae$ and $\te$ constraint and the dotted curve is the distribution 
obtained with only the $\te$ constraint.
\smallskip
}
\label{fig:eight}
\end{figure}
% --------------------------------------------------------------

In the Bayesian analysis, we conduct a simulation of Galactic lensing events using the prior models 
of the mass function of astronomical objects in the Galaxy and their physical and dynamical distributions.  
For the mass function, we consider both stellar and remnant lenses, i.e., black holes, neutron stars, 
and white dwarfs, by adopting the \citet{Chabrier2003} model and the \citet{Gould2000} model for the mass 
functions of stars and remnants, respectively.  In the simulation, lenses and source are located following 
the physical distribution model of \citet{Han2003b} and their motions are computed using the dynamical model 
of \citet{Han1995}.  We produce $10^7$ artificial lensing events, from which the probability distributions 
of $M$ and $\dl$ are obtained with the constraints of the measured $\te$ and $\mu$.

In Figure~\ref{fig:eight}, we present the probability distributions of the lens mass of the host star
($M_{\rm host}$, upper panel) and the lens-source separation ($D_{\rm LS}$, lower panel) obtained from 
the Bayesian analysis. As indicated by the small angular Einstein radius, the lens is estimated to lie 
close to the source, and thus we present the distribution of $D_{\rm LS}$ rather than $\dl$. To check 
the importance of the $\mu$ constraint, we present two sets of distributions obtained with the 
combined $\mu$ and $\te$ constraint (solid curves) and with only the $\te$ constraint (dotted curves). 
The distributions show that the lens mass estimated with the additional $\mu$ constraint is substantially 
lower and the lens-source separation is smaller than those estimated with the single $\te$ constraint. This 
indicates that the measured $\mu$ provides an important constraint on the physical lens parameters.

In Table~\ref{table:four}, we list the estimated physical lens parameters.  We note that both the 
inner and outer 2L1S solutions result in similar parameters, and thus we present the parameters based 
on the inner 2L1S solution.  The presented parameters are the median values of the Bayesian 
distributions, and the upper
and lower limits correspond to the 15.9\% and 84.1\% of the distributions.  It is found 
that the lens is a planetary system composed of a giant planet and a low-mass host star. The masses of 
the planet and host are
\begin{equation}
M_{\rm p} = 0.79^{+1.30}_{-0.39}~M_{\rm J}
\label{eq9}
\end{equation}
and
\begin{equation}
M_{\rm host} = 0.10^{+0.17}_{-0.05}~M_\odot,
\label{eq10}
\end{equation}
respectively. The planetary system is located in the bulge with a line-of-sight separation
from the source star of
\begin{equation}
D_{\rm LS}= 1.21^{+0.96}_{-0.63}~{\rm kpc}.
\label{eq11}
\end{equation}
The planet and host are separated in projection by
\begin{equation}
a_\perp= 
3.4^{+2.1}_{-1.0}~{\rm au}.
\label{eq12}
\end{equation}
Considering that the snowline of the system is $a_{\rm sl}\sim 2.7~{\rm au}(M_{\rm host}/M_\odot)\sim 0.4$~au, 
the planet is a wide-separation planet located well beyond the snowline of the host star.

% Table 4 ------------------------------------------------
\begin{deluxetable}{lcc}
\tablecaption{Physical lens parameters\label{table:four}}
\tablewidth{240pt}
%\tabletypesize{\small}
\tablehead{
\multicolumn{1}{c}{Parameter}          &
\multicolumn{2}{c}{Constraint}         \\
\multicolumn{1}{c}{}                   &
\multicolumn{1}{c}{$\te$ + $\thetae$}  &
\multicolumn{1}{c}{$\te$ only}           
}
\startdata                                              
$M_{\rm p}$ ($M_{\rm J}$)  &   $0.79^{+1.30}_{-0.39}$   &  $4.98^{+3.05}_{-2.94}$  \\
$M_{\rm host}$ ($M_\odot$) &   $0.10^{+0.17}_{-0.05}$   &  $0.68^{+0.42}_{-0.41}$  \\
$D_{\rm LS}$ (kpc)         &   $1.21^{+0.96}_{-0.63}$   &  $2.60^{+1.29}_{-1.13}$  \\
$a_\perp$ (au)             &   $3.4^{+2.1}_{-1.0}$      &  $11.5^{+3.1}_{-4.3}$
\enddata                            
\tablecomments{
The presented parameters are the median values of the Bayesian
distributions, and the upper and lower limits correspond to the
15.9\% and 84.1\% of the distributions.
\bigskip
}
\end{deluxetable}
\bigskip
% --------------------------------------------------------------

\section{Discussion and Conclusion}\label{sec:seven}

We analyzed the microlensing event OGLE-2016-BLG-1227, for which the event timescale 
was short and the light curve was affected by severe finite-source effects.  The light 
curve appeared to be that of a 1L1S event and the analysis based on the 1L1S interpretation 
yielded a short timescale and a very small angular Einstein radius, suggesting that the 
lens could be an FFP.  From the close inspection of the small residuals from the 1L1S 
solution, we found that the residual was explained by the existence of an additional 
widely separated heavier lens component, indicating that the lens was a planetary system 
with a wide-separation planet rather than an FFP.  From the Bayesian analysis with the 
constraints of the measured event timescale and relative lens-source proper motion, we estimated 
that the lens was composed of a planet with a mass 
$M_{\rm p} = 0.79^{+1.30}_{-0.39}~M_{\rm J}$ 
and a host star with a mass 
$M_{\rm host}=0.10^{+0.17}_{-0.05}~M_\odot$.  
It turned out that the planet was located well beyond the snowline of the host with a projected 
separation of $a_\perp= 3.4^{+2.1}_{-1.0}~{\rm au}$ It was estimated that the lens was located 
close to the source with a lens-source separation of $D_{\rm LS}=1.21^{+0.96}_{-0.63}$~au.

The event demonstrates that detecting deviations from 1L1S light curves provides an 
important method to distinguish wide-separation planets from FFPs.  Besides OGLE-2016-BLG-1227, 
there were two planetary events, in which planets were detected through isolated events and 
their widely separated hosts were identified in lensing light curves.  The first case  is 
MOA-bin-1 \citep{Bennett2012}.  For this event, the lensing light curve exhibited 
little lensing magnification attributable to the host of the planet similar to OGLE-2016-BLG-1227, 
but the planetary signal was entirely due to a brief caustic feature.  The second case is 
OGLE-2008-BLG-092 \citep{Poleski2014}. For this event, the planet was detected through the 
isolated event channel, but in this case the host of the planet was on the source trajectory, 
and gave rise to a bump in the lensing light curve.  OGLE-2016-BLG-1227 shows that there are 
a range of deviations in the signatures of host stars and that it is possible to identify the 
existence of a host even when a deviation is subtle.

Due to the unusual nature of OGLE-2016-BLG-1227, in which the relative lens-source proper 
motion $\mu=\thetae/\te$ is well determined, but the separate values of $\thetae$ and 
$\te$ are poorly constrained, the information that can be obtained from high-resolution 
follow-up observations would be different from that of normal events.  If follow-up observations 
are conducted to normal events with well estimated $\thetae$, the flux from the host is measured 
and from this one can make a diagram of the predicted host flux in $M$--$\dl$ plane.  Comparison 
of this diagram to $\thetae$ constraint in the same $M$--$\dl$ plane will allow one to determine 
$M$ and $\dl$ from the intersection of these two constraints, e.g., \citet{Yee2015} and 
\citet{Fukui2019}.  Even if $\thetae$ is not known because of poor $\rho$ measurement, the 
event timescale $\te$ is known.  Then, from late time follow-up imaging conducted when the 
source and lens are separated, one can measure the lens-source separation $\Delta\theta$ and 
therefore the relative lens-source proper motion can be estimated by $\mu = \Delta\theta/\Delta t$, 
from which the angular Einstein radius is estimated by $\thetae=\mu\te$ .  Here $\Delta t$ 
represents the difference between the time of follow-up observation and $t_0$.

For events with a well measured $\mu$ but with uncertain values of $\thetae$ and $\te$, the 
time of follow-up observations can be predicted.  If follow-up observation is conducted using 
the European Extremely Large Telescope (E-ELT) with an aperture of 39~m, the full width half 
maxima (FWHM) in the $J$ and $H$ band would be  ${\rm FWHM}(J)\sim 7.1$~mas and 
${\rm FWHM}(H)\sim 10.3$~mas, respectively.  Assuming that the lens and source can be resolved 
when they are separated by $\sim 1.5\times {\rm FWHM}$, the required times for the resolution 
would be $\Delta t\sim 13.5$ years and $\sim 19.6$~years from $J$ and $H$ imaging observations, 
respectively.  These correspond to the years 2028 and 2035, respectively.  With a resolved host 
star, its distance $\dl$ and mass $M_{\rm host}$ would be constrained from the color and flux.

However, this does not necessarily imply that the planet mass $M_{\rm p} = M_{\rm host}/q$ 
can also be well determined because the mass ratio is poorly known.  If one can estimate 
$M_{\rm host}$ and $\dl$ from the $J$ and $H$ color and magnitude, then there will two possible 
cases.  If the lens is in the disk, one can estimate 
$\pi_{\rm rel} = {\rm au}(D_{\rm L}^{-1} - D_{\rm S}^{-1})$, where $\ds\sim 9$~kpc.  Then the 
Einstein radius can be determined by the relation in Equation~(\ref{eq1}), although uncertainty 
will be fairly large because $M_{\rm host}$ and $\dl$ are somewhat uncertain together with the 
uncertainty of the source distance.  If the lens is in the bulge, in contrast, it will be difficult 
to estimate $\thetae$ any better than from the microlensing data.
This will cause $q$ and $M_{\rm p}$ to be poorly constrained.

\acknowledgments
Work by CH was supported by the grants  of National Research Foundation of Korea 
(2017R1A4A1015178 and 2019R1A2C2085965).
% Gould  
Work by AG was supported by US NSF grant AST-1516842 and by JPL grant 1500811.
AG received support from the European Research Council under the European Union's 
Seventh Framework Programme (FP 7) ERC Grant Agreement n.~[32103].
% OGLE  
The OGLE project has received funding from the National Science Centre, Poland, grant
MAESTRO 2014/14/A/ST9/00121 to AU.
% KMTNet
This research has made use of the KMTNet system operated by the Korea
Astronomy and Space Science Institute (KASI) and the data were obtained at
three host sites of CTIO in Chile, SAAO in South Africa, and SSO in
Australia.
% KREONET
We acknowledge the high-speed internet service (KREONET)
provided by Korea Institute of Science and Technology Information (KISTI).

\end{document}